\documentclass[fleqn,twoside]{article}
\usepackage{espcrc2}
\usepackage{graphicx}
\usepackage[figuresright]{rotating}

\title{
Megaton Water Cerenkov Detectors and Astrophysical Neutrinos}
\author{M. Goodman\address[HEP]{High Energy Physics Division
\\HEP 362 \\ Argonne Illinois 60439}}
\begin{document}
\begin{abstract}
Although formal proposals have not yet been made, the UNO and
Hyper-Kamiokande projects are being developed to follow-up the
tremendously successful program at Super-Kamiokande using a detector
that is 20-50 times larger.  The potential of such a detector to
continue the study of astrophysical neutrinos is considered and
contrasted with the program for cubic kilometer neutrino observatories.
\end{abstract}
\maketitle
\section{Introduction}
There are a variety of possible sources for astrophysical neutrinos
in the Universe, including neutrinos from recent supernovae, 
supernova relic neutrinos, decays of exotic particles,
beam dumps associated with active galactic nuclei, the galactic disk, and many 
other hypothetical sources.  Based on the successes of the Super-Kamiokande
project which was designed to search for proton decay and study other 
neutrino processes, plans are being formulated for larger water Cerenkov
detectors to extend our reach in proton lifetime and neutrino flux.  The word
``megaton" has possibly negative connotations with regard to nuclear weapons,
so I will refer to these new projects as ``thousand milli-Megaton" detectors.
\par In the next section I will describe a search for AGN neutrinos using the
1 milli-Megaton Soudan~2 detector.  This is neither new nor a water Cerenkov
detector, but it has not previously been described at a NOW workshop.
In the following sections, I will consider the capabilities of
UNO and/or Hyper-Kamiokande to astrophysical neutrinos.
The capabilities for solar neutrinos and galactic supernovae are left to
other sessions of this workshop.
\section{Search for AGN neutrinos in Soudan~2.}
The one milli-megaton Soudan~2 detector, built to search
for nucleon decay, was located approximately 700 meters
below the Earth's surface in an historic iron mine in Soudan, Minnesota.  It
was a very-fine grained detector, with drift tubes between thin steel plates,
and read out by wire chambers and cathode pads.  
A muon traversing the detector would go through a few hundred gas crossings,
with pulse height readout for each hit.
The overburden and rock composition were such that atmospheric muons died out
at approximately 82 degrees zenith angle, though the actual cut was
applied using a rock density and topology map corresponding to 14 kmwe.
  Incoming muons with an angle larger
than this were candidate atmospheric neutrino induced muons.  Since Soudan~2
could not distinguish upward from downward going muons, this corresponded to
a solid angle for neutrino astronomy of 1.77 sr or 14\% of $4\pi$.  A cut on
multiple scattering in the detector of less than 2 degrees was applied in order
to reject vertical
muons which may have scattered before arriving at the detector.
With an effective area of 86.7 m$^2$, a trigger and reconstruction efficiency of 
0.53 and a live time of $2.0 \times 10^8$s, 65 neutrino-induced horizontal
muon candidates were detected, corresponding to a flux of $4.01 \pm 0.50 \pm 0.30 
\times 10^{-13} {\rm cm}^{-2} {\rm sr}^{-1} {\rm s}^{-1}$.  
\par
\begin{figure}[t]
\includegraphics[width=3in]{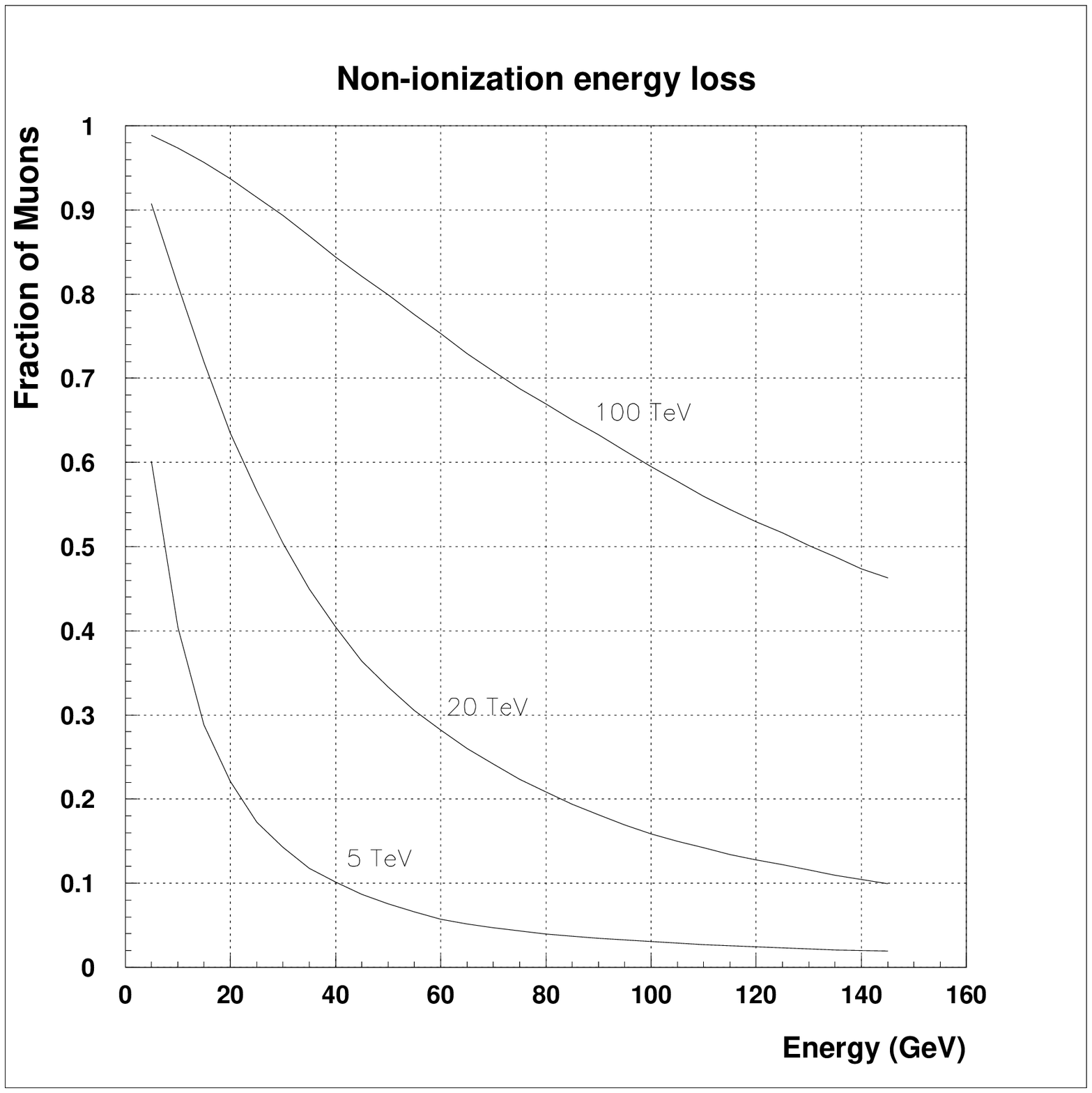} 
\caption{ 
Calculation of non-ionization energy loss for muons in Soudan~2}
\label{fig:eloss}
\end{figure}

 This data set was used to perform a search for diffuse production of
neutrinos by Active Galactic Nuclei.  Models for AGN neutrinos predict a
rather flat spectrum with substantial fluxes of neutrinos up to 100 TeV.
These neutrinos would then produce high energy muons with energies from 
a few TeV up to 100s of TeV.  In this energy range, the dominant energy loss 
process for the muons in iron is electron pair production followed by
bremsstrahlung.  A GEANT-based Monte Carlo was used to study energy
deposition in Soudan 2 by 5, 20 and 100 TeV muons.  The result is shown
in Figure \ref{fig:eloss}.  60\% of the 5 TeV muons lose 5 GeV or more
while traversing the Soudan~2 detector, along
with 91\% of 20 TeV muons and 99\% of 100 TeV muons.  Such an energy loss
is visible as a huge electromagnetic shower consisting of over 100 additional
hits along the trajectory of the muon.  None of the 65 muons had
visible radiated energy loss greater than the predetermined cut of 5 GeV.
This result allowed us to calculate an energy dependent flux limit for
neutrino-induced muons.  For 5, 20 and 100 TeV muons, the limits
are 2.2, 1.5 and 1.4 $\times 10^{-14}$ cm$^{-2}$sr$^{-1}$s$^{-1}$.

\section{Future Searches for High-Energy Astrophysical Neutrino Sources}
In order to search for neutrino-induced muons, the important property for
a detector is its area.  Super-Kamiokande is a 50 milli-Megaton detector.
UNO as presently designed will be 13 times the volume of Super-K, but only
5.5 times the area, or 6600 m$^2$.  For a background dominated signal,
this will only improve limits by a factor of $\sqrt{5.5}=2.3$.  In contrast
cubic kilometer detectors will have areas of about a million m$^2$ and
are already under construction.  It is unlikely that Megaton water
Cerenkov Detectors will be competitive for any signal which will be triggered by
the larger detectors.
Possible neutrino signals which fall in this category include the following:
\begin{itemize}
\item Searches for diffuse neutrino production from Active Galactic Nuclei
\item $\nu_\mu$ from WIMP annihilation products from the center of the earth
\item $\nu_\mu$ from WIMP annihilation products from the sun
\item Searches for WIMP annihilation in the galactic core
\item Searches for point sources of VHE/UHE $\nu_\mu$
\item Search for $\nu_\mu$ from the galactic disk (aka galactic atmospheric
neutrinos)
\item Search for coincidences with gamma ray bursts and
other astrophysical phenomena
\end{itemize}

\par The search for point sources provides a particular challenge, because
there are a variety of techniques to characterize the background.  While
the angular resolution of the detector can be known in principle, the
energy distribution of the hypothetical source affects the analysis strategy
for choosing the angular range.  All-sky surveys have been done using bins, cones
and unbinned searches.  Each has advantages and disadvantages for particular
kinds of sources.  For examples cones have fewer bin-edge problems,
but there are odd oversampling effects to deal with.
The use of multiple tests leads to multiple trial factors and hence a
lower true sensitivity for a real signal.
\par As pointed out above, these signals typically depend on the area of 
the detector, so the cubic kilometer arrays will have a clear advantage
over Megaton water Cerenkov detectors.  However, as the search for
AGN $\nu$s with Soudan~2 showed, advantage can be taken of the additional
information that a finer grained detector provides.  The signals may show up
only after a series of cuts that can only be made using the greater information
in the finer grained detectors.  Or a signal from the sun may be difficult if
the only cubic kilometer detector is at the south pole where the sun is often
near the horizon.  Astrophysical sources may not be a primary motivation for
Megaton Cerenkov detectors, but when built, searches should be made!
\section{Supernova Relic neutrinos}
Supernova relic neutrinos (SNR) is the name given to the diffuse
sum of all supernovae at cosmological distances from long long ago
and far far away.  The energy spectrum
for these neutrinos will be lower
than the distribution from a galactic supernova due to the red shift.
The observation strategy is to look between 18 MeV which is the upper
end of the hep solar $\nu$ flux and about 30 MeV where atmospheric
neutrinos will dominate.  Observation of SNR $\nu$s will provide a
direct test of various early star-formation models.  A variety
of authors\cite{bib:snr} have made predictions for the flux ranging from
$0.30$ to $1.2~{\rm cm}^{-2}{\rm s}^{-1}$ based on the ultraviolet luminosity and other
assumptions.
\par
\begin{figure}[htb]
\includegraphics[width=3in]{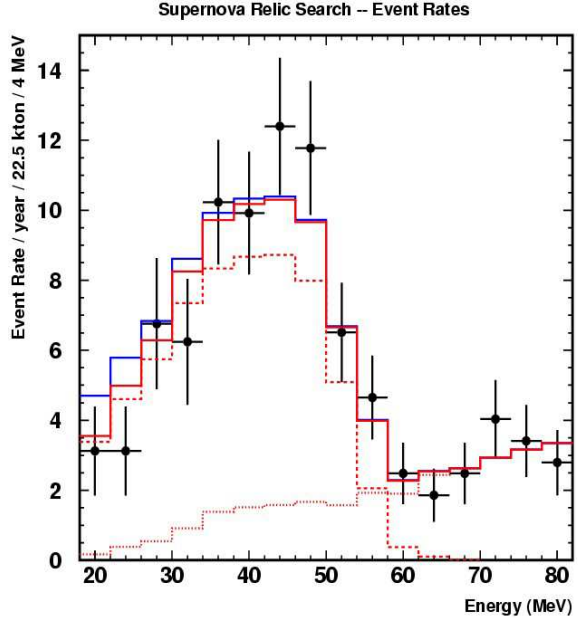}
\caption{ A fit to the observed spectrum of low energy electrons
in Super-K, from Reference \cite{bib:sksnr}.}
\label{fig:sksnr}
\end{figure}

 The Super-Kamiokande collaboration has conducted a search for
SNR neutrinos and set a limit at $1.2~{\rm cm}^{-2}{\rm s}^{-1}$.\cite{bib:sksnr} Their fit, which
is dominated by background from Michel electrons is shown in 
Figure~\ref{fig:sksnr}.
\par In UNO, with a planned 450 kiloton fiducial volume, a signal of 20-60 events
per year would be expected, which is enough to rise above the
backgrounds.  This is a background limited search, so the significance will
increase with the square-root of the exposure.  Due to spallation backgrounds,
this search would prefer a deeper location for UNO than might be required
for proton decay or long-baseline accelerator searches.  Depending on
the flux, a one sigma ``hint"
for  signal is expected in 0.5 to 6 years.

\section{Prospects for Megaton Water Cerenkov Detectors}
Super-Kamiokande has demonstrated the tremendous versatility of large
water Cerenkov detectors, by obtaining important results in the fields
of solar neutrinos, atmospheric neutrinos, accelerator neutrinos (K2K)
and supernova neutrinos, as
well as setting important limits on proton decay.  In my opinion, a
crucial reason to consider another generation of water detectors is to
continue the search for nucleon decay.  In most GUTs, there is an
{\it upper} limit for the lifetime of proton decay around a few $10^{36}$ years.
With current limits around $10^{34}$ years, most of the remaining parameter
space is accessible using Megaton detectors for ten years.  The UNO
collaboration had done a detailed study of backgrounds in such a detector,
and shown how the sensitivity increases with time.\cite{bib:unowhite}
\par The baseline design for UNO is for three volumes of water, each
$60\times60\times60~{\rm m}^3$.  The central volume would have 40\%
phototube coverage while the other two would have 10\%.  The fiducial
volume of 440 kton would be twenty times larger than Super-K.  There would
be 56,000 20 inch PMTs and 14,900 8 inch PMTs.  Possible sites for
UNO include the FREJUS tunnel in France, and a number of sites in the U.S.
The Frejus site is being considered in conjunction with a new safety tunnel
to be constructed alongside the existing highway tunnel.  In the U.S.,
a process has been identified to consider sites for a Deep Underground
Science and Engineering Laboratory (DUSEL) by the National Science Foundation.
Sites under consideration include the Kimballton site in Virginia, the
Sudbury site in Canada, the Soudan site in Minnesota, the Homestake site
in South Dakota, the WIPP site in New Mexico, the Henderson site in Colorado,
the San Jacinto site in California and the Icicle Creek site in Washington.
The collaboration has been working most closely with the Henderson mine,
which is owned by Climax Molybdenum Company, a subsidiary of Phelps Dodge
Corporation.  The mine produces 21,000 tons of ore per day, and has a lifetime
of another 20 years.  It is the 6th-7th largest underground hard rock mine
in the world.  A 28 foot diameter shaft from the surface can haul up to 200
people down to the 7500 foot level in about 5 minutes.
\par In Japan, a site for a one Megaton Detector, called Hyper-Kamiokande
has been identified at Tochibora, a few miles from the present site.
The new beam for T2K is aimed at the same angle with respect to the current
Super-K detector and the possible new detector Hyper-K.  
\par The Hyper-K idea is not presently before
the Japanese funding agencies.  The UNO collaboration provided its white
paper on its capabilities for the 2001 Snowmass meeting.   A new
proton decay experiment was identified by HEPAP on its wish list for
new long-term projects in March 2003.  The report, titled 
``High energy physics facilities recommended for the DOE office 
of science twenty-year roadmap", rates the scientific potential 
of UNO (Classified as Underground Detector) as ``Absolutely Central".\cite{bib:hepap} 
However, this was the only absolutely central experiment which did not make it
onto the smaller list of projects in the DOE Office of Science's 20 year
strategic plan.\cite{bib:doeplan}
The APS neutrino study, in its ``Neutrino Matrix" report stated
its assumption that adequate R\&D would be carried out ``to assure
the practical and timely realization of accelerator and detector
technologies critical to the recommended program.  Of particular
importance are R\&D efforts aimed toward development of ... a very
large neutrino detector..."  Megaton water Cerenkov detectors, with a price
tag approaching a billion dollars, are probably not on the threshold of
being approved.  However, the case for such detectors continues to solidify,
and I hope they will become a reality in the not-too-distant future.
\section{Acknowledgments}  I am grateful to the UNO collaboration 
and in particular to Alec Habig for information about Super-Kamiokande
searches and consideration of UNO capabilities.

\end{document}